\documentclass[12pt,a4paper,twoside]{paper}
\usepackage{amsmath,bbm,amssymb,amsxtra}
\usepackage{graphicx}
\usepackage{epstopdf}
\usepackage{chngcntr}
\usepackage{etoolbox}

\usepackage{bbold}

\usepackage{fancyhdr}
 \usepackage{fancyhdr}
\newcommand{\be}{\begin{equation}}
\newcommand{\ee}{\end{equation}}

\newcommand{\ba}{\begin{eqnarray}}
\newcommand{\ea}{\end{eqnarray}}

\newcommand{\bea}{\begin{eqnarray*}}
\newcommand{\eea}{\end{eqnarray*}}

\newcommand{\bee}{\begin{enumerate}}
\newcommand{\ene}{\end{enumerate}}

\usepackage{epsfig}
\def\R{\mathbb R}

\usepackage{appendix}
\usepackage{chngcntr}
\usepackage{etoolbox}
\usepackage{lipsum}
\usepackage{amsmath,bbm,amssymb,amsxtra}
\usepackage{enumerate}
\allowdisplaybreaks
\usepackage[active]{srcltx}
\usepackage{epsfig}
\usepackage{ae}
\numberwithin{equation}{section}

\newtheorem{lemma}{Lemma}[section]
\newtheorem{proposition}[lemma]{Proposition}

\newtheorem{theorem}[lemma]{Theorem}

\newtheorem{remark}[lemma]{Remark}
\newtheorem{definition}[lemma]{Definition}

\newcommand{\prop}[1]{\begin{proposition}\label{#1}
\sl }
\newcommand{\eprop}{\end{proposition}}
\newcommand{\thm}[1]{\begin{theorem}\label{#1}
\Ä }
\newcommand{\ethm}{\end{theorem}}
\newcommand{\lem}[1]{\begin{lemma}\label{#1}
\sl }
\newcommand{\elem}{\end{lemma}}

\newcommand{\defin}[1]{\begin{definition}\label{#1}
\sl }
\newcommand{\edefin}{\end{definition}}

\def\la{\lambda}

\def\CA{{\mathcal A}}              
              
       \def\CH{{\mathcal H}}

\def\CH{{\mathcal H}}

\def\qq{ \begin{eqnarray} }
\def\qqq{ \end{eqnarray} }
\def\rr{ \begin{equation} }
\def\rrr{ \end{equation} }

\def\qq{ \begin{eqnarray} }
\def\qqq{ \end{eqnarray} }

\usepackage{epsfig}

\counterwithin*{equation}{section} %sd: pour avoir un numéro d'équation de type (9.9.9)
\counterwithin*{equation}{subsection} %sd: pour avoir un numéro d'équation de type (9.9.9.9)
\ifx\optionkeymacros\undefined\else \fi

\catcode`\Œ=\active\defŒ{{\aa}}       % option a
\catcode`\º=\active\defº{\int}        % option b (math mode) 
\catcode`\=\active\def{\c c}        % option c
\catcode`\¶=\active\def¶{\partial}    % option d (math mode)
\catcode`\Ä=\active\defÄ{\oint}       % option f (math mode) ?
\catcode`\Æ=\active\defÆ{\triangle}   % option j (math mode)
\catcode`\Â=\active\defÂ{\neg}        % option l (math mode)
\catcode`\µ=\active\defµ{\mu}         % option m (math mode)
\catcode`\¿=\active\def¿{{\o}}        % option o
\catcode`\¹=\active\def¹{\pi}         % option p (math mode w/ arg.)
\catcode`\Ï=\active\defÏ{{\oe}}       % option q 
\catcode`\§=\active\def§{{\ss}}       % option s 
\catcode`\ =\active\def {\dagger}     % option t  (math mode)
\catcode`\Ã=\active\defÃ{\sqrt}       % option v (math mode w/ arg.)
\catcode`\·=\active\def·{\Sigma}      % option w (math mode)
\catcode`\Å=\active\defÅ{\approx}     % option x (math mode)
\catcode`\½=\active\def½{\Omega}      % option z (math mode)
\catcode`\£=\active\def£{{\it\$}}     % option 3 ($ from italic font)
\catcode`\°=\active\def°{\infty}      % option 5 (math mode)
\catcode`\¤=\active\def¤{{\S}}        % option 6 
\catcode`\¦=\active\def¦{{\P}}        % option 7
\catcode`\¥=\active\def¥{\bullet}     % option 8 
\catcode`\»=\active\def»{\leavevmode\raise.585ex\hbox{\b a}}      % option 9
\catcode`\¼=\active\def¼{\leavevmode\raise.6ex\hbox{\b o}}        % option 0
\catcode`\­=\active\def­{\not=}       % option = (math mode)
\catcode`\²=\active\def²{\leq}        % option , (math mode)
\catcode`\³=\active\def³{\geq}        % option . (math mode)
\catcode`\Ö=\active\defÖ{\div}        % option / (math mode)
\catcode`\É=\active\defÉ{{\dots}}     % option ; 
\catcode`\¾=\active\def¾{{\ae}}       % option '
\catcode`\Ç=\active\defÇ{\ll}         % option \ (math mode)
\catcode`\Ò=\active\defÒ{``}          % option [
\catcode`\Á=\active\defÁ{!`}          % option !
\catcode`\¢=\active\def¢{\rlap/c}     % option 4
\catcode`\Ô=\active\defÔ{`}           % option ] 
\catcode`\Õ=\active\defÕ{'}           % shift option ]

\catcode`\=\active\def{{\AA}}       % shift-option A
\catcode`\'=\active\def'{\c C}        % shift-option C
\catcode`\¯=\active\def¯{{\O}}        % shift-option O
\catcode`\¸=\active\def¸{\Pi}         % shift-option P (math mode)
\catcode`\Î=\active\defÎ{{\OE}}       % shift-option Q
\catcode`\®=\active\def®{{\AE}}       % shift-option '
\catcode`\×=\active\def×{\diamond}    % shift-option V (math mode)
\catcode`\¡=\active\def¡{\accent'27}  % shift-option 8
\catcode`\Ó=\active\defÓ{''}          % shift-option [
\catcode`\±=\active\def±{\pm}         % shift-option = (math mode)
\catcode`\È=\active\defÈ{\gg}         % shift-option \ (math mode)
\catcode`\À=\active\defÀ{?`}          % shift-option / 
\catcode`\Ð=\active\defÐ{--}          % option - (en-dash)
\catcode`\Ñ=\active\defÑ{---}         % shift-option - (em-dash)

\catcode`\Š=\active\defŠ{\"a}        % option u, then  a
\catcode`\'=\active\def'{\"e}        % option u, then  e
\catcode`\•=\active\def•{\"{\i}}     % option u, then  i
\catcode`\š=\active\defš{\"o}        % option u, then  o
\catcode`\Ÿ=\active\defŸ{\"u}        % option u, then  u
\catcode`\Ø=\active\defØ{\"y}        % option u, then  y
\catcode`\€=\active\def€{\"A}        % option u, then  A
\catcode`\…=\active\def…{\"O}        % option u, then  O
\catcode`\†=\active\def†{\"U}        % option u, then  U
\catcode`\‡=\active\def‡{\'a}        % option e, then  a
\catcode`\Ž=\active\defŽ{\'e}        % option e, then  e
\catcode`\'=\active\def'{\'{\i}}     % option e, then  i
\catcode`\—=\active\def—{\'o}        % option e, then  o
\catcode`\œ=\active\defœ{\'u}        % option e, then  u
\catcode`\ƒ=\active\defƒ{\'E}        % option e, then  E
\catcode`\ˆ=\active\defˆ{\`a}        % option `, then  a
\catcode`\=\active\def{\`e}        % option `, then  e
\catcode`\"=\active\def"{\`{\i}}     % option `, then  i
\catcode`\˜=\active\def˜{\`o}        % option `, then  o
\catcode`\=\active\def{\`u}        % option `, then  u
\catcode`\Ë=\active\defË{\`A}        % option `, then  A
\catcode`\‹=\active\def‹{\~a}        % option n, then  a
\catcode`\–=\active\def–{\~n}        % option n, then  n
\catcode`\›=\active\def›{\~o}        % option n, then  o
\catcode`\Ì=\active\defÌ{\~A}        % option n, then  A
\catcode`\"=\active\def"{\~N}        % option n, then  N
\catcode`\Í=\active\defÍ{\~O}        % option n, then  O
\catcode`\‰=\active\def‰{\^a}        % option i, then  a
\catcode`\=\active\def{\^e}        % option i, then  e
\catcode`\"=\active\def"{\^{\i}}     % option i, then  i
\catcode`\™=\active\def™{\^o}        % option i, then  o
\catcode`\ž=\active\defž{\^u}        % option i, then  u

\let\optionkeymacros\null

\begin{document}

%\vspace*{10mm}

     \begin{center}
 \textsl{\LARGE  From EPR, Schr\"odinger paradox to nonlocality based on perfect correlation
  }

 \vspace*{5mm}
 { \Large Jean Bricmont
 
 IRMP,
Universit\'e catholique de Louvain,
chemin du Cyclotron 2,
1348 Louvain-la-Neuve
Belgium}\footnote{E-mail: jean.bricmont@uclouvain.be}
 \vspace*{5mm}
 
{ \Large Sheldon Goldstein

 Department of Mathematics, Rutgers University, Hill Center, 110 Frelinghuysen Road, Piscataway,
NJ 08854-8019, USA}\footnote{E-mail: oldstein@math.rutgers.edu}

 \vspace*{5mm}
{ \Large Douglas Hemmick

 66 Boston Drive,
Berlin, MD 21811, USA}\footnote{E-mail: jsbell.ontarget@gmail.com}

\end{center}

\begin{abstract}

We give a conceptually simple proof of nonlocality (based on the previous work of \cite{Hem, Hem-S, 1, 2}) using only the perfect correlations between results of measurements on distant systems  discussed by Einstein, Podolsky and Rosen---correlations that EPR  thought proved the incompleteness of quantum mechanics. 
 Our argument relies on an extension of EPR by Schr\"odinger.

\end{abstract}

\section{Introduction}\label{sec1}

The implications of the Einstein Podolsky Rosen paradox and Bell's
theorem have fascinated many physicists and physics students for
decades. While there is essentially universal agreement that something
important has been proven, exactly what that is is
controversial. There
are those\footnote{See \cite{Bethe},
\cite[p. 172]{Gell-Mann},  \cite{Mermin early AJP} and
\cite{Wigner Big Red}.}
who take the position that the ``hidden variables" whose existence was asserted by EPR
 do not exist, which in turn leads some authors to doubt the existence of ``objective reality."\footnote{The most extreme position might be that of
 Mermin, who told us that (due to a variant of Theorem \ref{2} below) 
the moon is demonstrably not there
 when nobody looks \cite[p. 397]{Mermin}.} 

In
concurrence with 
other authors,\footnote{See, for example, \cite{Albert}, \cite{Bell cascade photons}, \cite{Bertlemann's Socks}, \cite{Intuitive Nonlocality}, \cite[Section 
  8]{ShellyRecent}, 
\cite{Baby Snakes}, \cite{TimLaud}, \cite{Ma},  \cite{Boxes},
\cite{locality},  \cite{Shadows} and \cite{Wiseman}.}
we claim that it is rather nonlocality which is the inescapable conclusion of the combinations of the arguments of EPR and of Bell. However, the originality of our analysis is to use an observation by Schr\"odinger \cite{Sch}  implying that the perfect correlations alone suffice to establish nonlocality. Thus, no additional relationship on the observables, e.g., Bell's
inequalities, must be empirically tested to establish that conclusion.

Many authors\footnote{ See for example, \cite{Laloe}
\cite{KS}, \cite{Mermin's Gleason} and \cite{Reality Marketplace}.
}
 argue correctly that a consequence of the theorems on the non-existence of ``hidden variables" (such as Theorem \ref{2} below or variants of it) is that quantum
observables cannot have  values pre-existing their measurement. But few\footnote{
See, for example,  \cite{Be1}, \cite{Bell Imposs Pilot} and \cite{Travis Spin}.
}
 explain how that is nonetheless
compatible with the existence and  success of what is arguably the most natural hidden variable approach to quantum mechanics--and indeed the most natural and obvious version of quantum mechanics itself--namely Bohmian mechanics 
(see Section  \ref{sec7}). 

To explain our reasoning,
let us start with a physically classical situation: consider the proverbial Alice and Bob, situated far away from each other, and simultaneoulsy tossing coins, over and over. One would expect the results on both sides to be random and uncorrelated. But suppose that the results appear indeed random but are also perfectly correlated: each time Alice's toss results in heads, Bob's toss also results in heads and similarly for tails.

Obviously such a strange situation would cry out for an explanation. One possibility is the following.
First, Alice and Bob are able to manipulate their coin tosses so as to obtain whichever results they desire and second,
they agree in advance on an apparently random sequence of results and manipulate their coin tosses so as to both obtain that sequence when they toss their coins.

This looks extravagant, but is there any other possibility? Well, yes, there exists an even more extravagant one: that when Alice tosses her coin, she instantly affects the trajectory of Bob's coin, so that Bob's coin falls on the same side as Alice's coin. 

Of course, this looks even more incredible than the previous scenario. But we may still ask: is there a third possibility? We don't see any and we will assume from now on that the reader agrees with us on that point.

In 1935, Einstein, Podolsky and Rosen (EPR) \cite{EPR} described a situation which is quite similar to the one above. In a reformulation of their argument due to David Bohm (\cite{Bo}), one considers two spin one-half particles moving in opposite spatial directions, in the spin state 
\be
|\Psi\rangle = \frac{1}{\sqrt 2} \big(|  \uparrow  \rangle| \downarrow  \rangle-|  \downarrow  \rangle|   \uparrow  \rangle\big),
\label{1}
\ee
% $|  \uparrow  \rangle$ means spin up  and $| \downarrow  \rangle$ means spin down.The state (\ref{1}) 
 where the right factors refer to particle $1$ and  left ones  to particle $2$.
 This state has the same form for all spin directions, and has the property that the result of the measurement of the spin of one particle 
in any given direction is perfectly anti-correlated with the result of the measurement of the spin of the other particle 
in the same direction, no matter how far apart the particles are: if the result of the measurement of the spin of  particle $1$ is ``up", the result for particle $2$ will be ``down" and vice-versa.

This raises the same question as the one about coin tosses: how does one explain that? But here the first possibility is much more plausible, a priori, than the analogous one about the coin tosses of Alice and Bob. We must simply assume that each particle carries with it ``instructions" telling it how to react (``up" or ``down") when its spin is measured in any given direction, and that the two particles have opposite instructions.

The problem is that the quantum state  (\ref{1}) does not include such instructions for individual particles. It  says only that the probability of ``up" and ``down"  are both equal to $\frac{1}{2}$, in any direction, but with the results for both particles being perfectly anti-correlated.

Hence, said EPR, in one of the most misunderstood, yet simple, arguments in the history of physics, the quantum state is an incomplete description of physical reality. It does predict the correct statistics, but does not describe completely the physical state of individual systems.
 Stated more precisely, it says that we must describe this pair of particles 
not only by their joint quantum state but also by other variables, often called ``hidden,"  that determine the behavior of those particles when one measures their spin in a given direction.

What could be wrong with this conclusion? In 1964, John Bell  showed that simply assuming the existence of these variables leads to a contradiction with the quantum predictions for the results of measuring 
the spin of those particles {\it in different directions}, one for the first particle and another for the second one (see \cite{DGTZ} for a simple proof of this contradiction).
Those predictions have been amply verified after Bell's publication (see (\cite{GNTZ} for a review).

But what does this imply? That we have no choice but to accept the analogue of the second branch of the alternative proposed about the coin tosses of Alice and Bob: that the measurement of the spin on one side affects instantaneously (if the measurements on both sides are made simultaneously), in some way, the result on the other side. This is what is called nonlocality or ``action at a distance." We write ``in some way" because this argument does not indicate how that action takes place and provides only an indirect proof of its existence.

We will use here an extension of the EPR argument due to  Schr\"odinger  \cite{Sch, Sch1, Sch2} to provide a more direct proof of nonlocality. We will also briefly sketch how nonlocality occurs in Bohmian mechanics. We will use very little mathematics and refer to the literature for proofs.

\section{Schr\"odinger's maximally entangled states}\label{sec2}

Schr\"odinger's extension of EPR argument relies on the use of special quantum states, called maximally entangled, for pairs of physical systems that may be located far apart. These states have the property that, for each quantum observable of one of the systems, there is an associated observable of the other one such that the result of the measurement of that observable is perfectly correlated with the result of the measurement of the first one.
  
Consider a finite dimensional (complex) Hilbert space $\cal H$, of dimension $N$,   and orthonormal bases $\psi_n$ and $\phi_n$ in $\cal H$ (we will assume below that all bases are orthonormal).
A unit vector, or state, $\Psi$ in $\cal H \otimes \cal H$ is {\it maximally entangled} if it is of the form 
\be
\Psi= \frac{1}{\sqrt N}\sum_{n=1}^N    \psi_n \otimes \phi_n.
\label{ME}
\ee

 Since we are interested in quantum mechanics, we will associate, by convention, each space in the tensor product with a ``physical system," namely we will consider the set  $\{\phi_n\}_{n=1}^N$ as a basis of states for physical system $1$  and the set  $\{\psi_n\}_{n=1}^N$ as a basis of states for physical system $2$.
 
 These states have the following fundamental properties (see \cite{1} for a proof):
 
 \begin{itemize}
\item[1]

The representation (\ref{ME}) is basis independent in the sense that, if we choose a basis
$\{\phi'_n\}_{n=1}^N$ for system $1$, instead of $\{\phi_n\}_{n=1}^N$, there is a basis $\{\psi'_n\}_{n=1}^N$ for system $2$
such that $\Psi$ can be written as:
\be
\Psi= \frac{1}{\sqrt N}\sum_{n=1}^N    \psi'_n \otimes \phi'_n.
\label{ME'}
\ee

\item[2]

Given a maximally entangled state $\Psi$, we may associate
 to every operator of the form  ${\mathbb 1}  \otimes O$ (acting on system $1$) an operator of the form  $ \tilde O \otimes {\mathbb 1}$ (acting on system $2$)
such that, 
if $(\phi_n)_{n=1}^N$ are the eigenstates of $O$, with eigenvalues $\la_n$,
\be
 O \phi_n = \la_n \phi_n,
\label{A1}
\ee
the vectors $(\psi_n)_{n=1}^N$ are the eigenstates  of $\tilde O$, also with eigenvalues $\la_n$:
\be
\tilde O \psi_n = \la_n \psi_n,
\label{A2}
\ee
(when $\Psi$ is of the form (\ref{ME})).
\end{itemize}

\begin{remark}\label{r1}
A simple example of a maximally entangled state is given in (\ref{1}).
 In this situation, we simply have $\tilde O=-O$ and the correlations mentioned below become anti-correlations.

\end{remark}

Let us now see what this notion of maximally entangled state implies for quantum measurements.

Suppose that we have a pair of physical systems, whose states  belong to the same finite dimensional  Hilbert space $\cal H$ (like spin states). And suppose that the quantum state $\Psi$ of the pair is maximally entangled, i.e. of the form (\ref{ME}).

Any observable acting on system $1$ is represented by a self-adjoint operator $O$, which has therefore a basis of eigenvectors. Since the representation (\ref{ME}) holds in any basis (for an appropriate choice of the basis $\{\psi_n\}_{n=1}^N$), let the set $\{\phi_n\}_{n=1}^N$  in (\ref{ME}) be the eigenstates of $O$ and let  $\la_n$  be the corresponding  eigenvalues, see  (\ref{A1}). 

 If one measures that observable $O$, the result will be one of the eigenvalues $\la_n$, each having equal probability $\frac{1}{N}$. If the result is $\la_k$,  the (collapsed) state of the system after the measurement will be $\psi_k \otimes \phi_k $. Then, the measurement of observable $\tilde O$ (with eigenstates $\psi_n$), on system $2$, will necessarily yield the value $\la_k$. 
 
Reciprocally, if one measures an observable $\tilde O$ on system $2$ and the result is  $\la_l$,  the (collapsed) state of the system after the measurement will be $\psi_l \otimes \phi_l $, and the measurement of observable $ O$ on system $1$ will necessarily yield the value $\la_l$. 

To summarize, we have derived the following consequence of the quantum formalism:

{\bf Principle of Perfect Correlations.}
{\it In any maximally entangled  quantum state, see (\ref{ME}), there is, for each operator $O$ acting on system $1$, an operator $ \tilde O$ acting on system $2$, such that, if one measures the physical quantity represented by operator $ \tilde O$ on system $2$ and the result is the eigenvalue  $\la_l$ of $ \tilde O$, then measuring the physical quantity represented by operator 
$O$  on system $1$ will yield with certainty the same eigenvalue $\la_l$, and vice-versa.}%\footnote{The correlations mentioned here are often called  anti-correlations, e. g. in the example of the spin in Remark \ref{r1} above, with $\tilde O=-O$.}}

\section{Schr\"odinger's ``Theorem"}\label{sec3}

The following property  will be  crucial in the rest of the paper.
 
{\bf Locality.}
{\it If systems $1$ and $2$ are spatially separated from each other, then measuring an observable on system $1$ has no instantaneous effect whatsoever on system $2$ and  measuring an observable on system $2$ has no instantaneous effect whatsoever on system $1$.}

Finally, we must define: 
 
 {\bf Non-contextual value-maps.}
  Suppose there are situations in which
the result of measuring an observable $A$ of a quantum system is determined already, before the measurement. Suppose, that is, that $A$ has, in these situations, a pre-existing value $v(A)$ revealed by measurement and not merely created by measurement. And suppose  that there is a situation  in which we have a pre-existing value $v(A)$ for every quantum observable $A\in \CA$, the set of self-adjoint operators on the  Hilbert space $\CH$ of the system.
  
 We would then have a 
 {\it  non-contextual value-map}, namely  a map  $v: \CA \to \R$ that assigns the value $v(A)$ to  any experiment associated with what is called in quantum mechanics a measurement of an observable $A$. There can be different ways to measure the same observable. The  value-map is called non-contextual  because all such experiments, associated with the same quantum observable $A$, are assigned the same value.

 A  non-contextual value-map has the fundamental property that if $A_i$, $i=1, \dots, n$, are mutually commuting self-adjoint operators on $\CH$, $
 [A_i, A_j]= 0, \forall i, j =1, \dots, n$, then, if $f$ is a  function of $n$ variables and $B= f(A_1, \dots, A_n)$,
 
 \be
v(B)= f(v(A_1), \dots, v(A_n)).
\label{res}
\ee
It is a well-known property of quantum mechanics that, since all the operators $A_1, \dots, A_n, B$ commute, they are simultaneously measurable with results that must satisfy (\ref{res}).

But, and this is important to  emphasize,  (\ref{res}) follows trivially from the non-contextualilty of the value-map. Indeed, a valid quantum mechanical  way to measure the operator $B= f(A_1, \dots, A_n)$ is to measure $A_1, \dots, A_n$ and, denoting the results $\la_1, \dots, \la_n$, to regard $\la_B=f(\la_1, \dots, \la_n)$ as the result of a measurement of $B$ .
Since, by the non-contextuality of the map $v$, all the possible measurements of $B$ must yield the same results, (\ref{res}) holds.

Now we will use the perfect correlations and locality to establish the existence of a non-contextual value-map $v$
for a maximally entangled  quantum state of the form (\ref{ME}). By the 
principle of perfect correlations,
for any  operator $O$ on system $1$, there is an  operator  $\tilde O$  on system $2$, which  is perfectly correlated with $O$ through (\ref{A1}, \ref{A2}).

Thus, if we were to measure  $\tilde O$, obtaining  $\la_l$, we would know that 
\ba
v(O)= \la_l 
\label{map}
\ea
concerning the result of then measuring $O$. Therefore, $v(O)$ would pre-exist the measurement of $O$.
But, by the assumption of locality, the measurement of   $\tilde O$, associated with the second system,  could not have 
had any effect on the first system,
and thus, this value $v(O)$ would pre-exist also  the measurement of  $\tilde O$ and this would not depend upon whether $\tilde O$ 
had been measured.
 Letting $O$ range over all operators on system $1$,  
we see that there must be a non-contextual value-map $O\to v(O)$. 

To summarize, we have shown:

 \noindent
{\bf Schr\"odinger's ``Theorem."} Let $\CA$ be the set of self-adjoint operators on the component Hilbert space   $\CH$ of a physical system 
in a maximally entangled state (\ref{ME}). Then, assuming locality and the principle of perfect correlations, there 
exists a non-contextual value-map  $v: \CA \to \R$.

\begin{remark}\label{r2}

We put ``Theorem" in quotation marks, here and below, when we refer to a non-contextual value-map, because its definition involves the physical notion of measurement, which is not  mathematically formalized. The  conclusions of this ``Theorem" are nevertheless inescapable  assuming the hypothesis of locality and the empirical validity of the principle of perfect correlations, a principle which is, as we saw, a direct consequence of the quantum formalism.
\end{remark}

\section{The non-existence of non-contextual value-maps}\label{sec4}

The problem posed by the non-contextual value-map $v$ whose existence is implied by Schr\"odinger's ``Theorem" is that such maps simply do not exist. Indeed, one has the:

\noindent
{\bf ``Theorem": Non-existence of non-contextual value-maps.}
Let $\CA$ be the set of self-adjoint operators on the  Hilbert space $\cal H$ of a physical system. Then there exists no non-contextual value-map $v: \CA \to \R$.
 
This ``Theorem" is an immediate consequence of the following purely mathematical result,
 since (\ref{res2}, \ref{res3}, \ref{res4}) are consequences of (\ref{res}) ((\ref{res3}, \ref{res4}) follow from (\ref{res}) by taking $n=2$ and $f(x,y)= x+y$ or $f(x,y)= xy$):

\begin{theorem}\label{2}
 Let $\cal H$ be  a finite dimensional  Hilbert space of dimension at least four,  and let $\CA$ be the set of self-adjoint operators on $\CH$.
There does not  exist a map  $v: \CA \to \R$ such that:

1) $ \; \forall O \in { \CA}$, 
\be
v(O) \;\;\mbox{is an eigenvalue of} \;\;O.
\label{res2}
\ee

2) Either
\be
v(O+ O')=v(O)+ v(O'),\label{res3}\\
\ee
$\forall O, O' \in { \CA}$ with  $ [O, O']= OO'-O'O=0$,
or
\be
v(O O')=v(O) v(O')\label{res4}\\
\ee
$\forall O, O' \in { \CA}$ with  $ [O, O']= OO'-O'O=0$,
holds.
\end{theorem}

See \cite{1} for a discussion of the proof of the Theorem, originally due to John Bell \cite{Be1} and to Kochen and Specker \cite{KS},  with simplified proofs due to David Mermin \cite{Me4}, and to Asher Peres \cite{Per, Per1}.

\section{Nonlocality}\label{sec5}

The conclusions of Schr\"odinger's ``Theorem" and of the ``Theorem"  on the non-existence of non-contextual value-maps plainly contradict each other.
So the assumptions of at least one of them must be false.  However, Theorem \ref{2} that implies the non-existence of non-contextual value-maps is a purely mathematical result. 
And  Schr\"odinger's ``Theorem" assumes only the perfect correlations and locality. The  perfect correlations are an immediate consequence of quantum mechanics.   The only remaining assumption  is locality. Hence we can deduce:

\noindent
{\bf Nonlocality ``Theorem"}. The predictions of quantum mechanics are incompatible with locality.

\begin{remark}\label{r3}

For a discussion of the relation between this proof  and other proofs  of nonlocality, including \cite{H-R, St, Brown, El, Cab, Ara}, see \cite[Sections 5, 7]{1}.

\end{remark}

\section{EPR's original argument}\label{sec6}

In their original paper \cite{EPR}, Einstein, Podolsky and Rosen considered the following formal maximally entangled state
 for two particles in one dimension: 
\begin{eqnarray}
\Psi_{EPR} (x_1, x_2)  &=& \int_{-\infty}^{\infty} \exp(i(x_1-x_2+x_0)p) dp.
\label{Psi_3}
\end{eqnarray}
(putting $\hbar=1$. We say formal because this state is not in the Hilbert space of the $2$-particle system).

Using a standard identity for distributions ($ \int_{-\infty}^{\infty} \exp(ixp) dp = 2\pi \delta (x)$) one can rewrite that state as:
 
\begin{eqnarray}
\Psi_{EPR} (x_1, x_2)  
&=& 2\pi \delta(x_1-x_2+x_0).  \nonumber\\
\label{Psi_4}
\end{eqnarray}

This state has, according to standard quantum mechanics, the property that, if one measures the position operator $Q_1$ of particle $1$ and obtains the result $x$, then the measurement of the position operator $Q_2$ of particle $2$ will yield  $x+x_0$, with a similar conclusion if one first measures $Q_2$. And, if one measures the momentum operator $P_1$ of particle $1$ and obtains the result $p$, then the measurement of the momentum  operator $P_2$ of particle $2$ will yield $-p$.

Using this fact, EPR claimed to have established that there are pre-existing values for $Q$ and $P$ (for are least one of the particles and in fact, by symmetry, for both) for a quantum system in the state (\ref{Psi_3}). For them, this proved the incompleteness of ordinary quantum mechanics, since it shows that certain quantities that are not part of the quantum formalism (the precise values of position and momentum) must exist (of course, here, they assumed  locality). 

Note that to obtain their conclusion of incompleteness it would have sufficed to consider just $Q$ (or just $P$)\footnote{This argument is discussed in detail by Maudlin \cite[3rd edn, pp.~128--132]{Ma}.}. In arguing for pre-existing values for both $Q$ and $P$ they established, assuming locality, the absolutely shocking conclusion that, despite the uncertainty principle, both position and momentum could have values at the same time. However, in so arguing, they somewhat obscured what they wished to establish: that the wave function of a quantum system does not provide its complete description, for which consideration of $Q$ alone would have sufficed.

\begin{remark}\label{r4}

What Schr\"odinger did was to recognize the full power of a maximally entangled state such as that of EPR, along  the lines described earlier in this paper. He did not realize that the pre-existing values so obtained were impossible. He did, however, realize that they were deeply puzzling, as follows: let $O$ be the energy of the harmonic oscillator, $O= \frac{1}{2} (
P^2+\omega^2 Q^2)$ with $P=-i \frac{d}{dx}$ (with $\hbar=1$). It is well known that the eigenvalues of the operator $O$ are of the form $\omega(n+\frac{1}{2})$, $n= 0,1, 2, \dots$. But, argued Schr\"odinger,  if these values $v(O)$ can be determined by measuring a similar operator $\tilde O$ acting on a distant system, they must pre-exist the measurement of $ O$, and that should hold true {\it for every value of $\omega$}. Similarly values $v(Q)$ and $v(P)$ of the position operator $Q$ and the momentum operator $P$ of the first system must also pre-exist their measurements. 

It would be natural to suppose, argued  Schr\"odinger, that $v(O)=\frac{1}{2} (v(P)^2+\omega^2 v(Q)^2)$. But values satisfying this relation for all $\omega$ are impossible: the quantities  $v(O)$ can't belong to the set $\{ \omega(n+\frac{1}{2}) | n= 0, 1, 2, \dots\}$, for all values of $\omega$ for any given values of
$v(Q)$ and $v(P)$. 

However, Schr\"odinger recognized that, since $Q$ and $P$ do not commute, the relation $v(O)=\frac{1}{2} (v(P)^2+\omega^2 v(Q)^2)$ need not hold and thus, that one cannot derive the above contradiction\footnote{Thus he did not make the mistake made by von Neumann (see \cite{2} for a discussion of that mistake).}.
Schr\"odinger concluded from the apparent contradiction that in fact $v(O)$ , $v(P)$, and $v(Q)$ can't be related in the same way as the operators $O$, $P$, and $Q$ are,  and he raised the question of exactly how the values are in fact related and indeed whether they are related at all: He raised the possibility that all these values, including $v(O)$ for all the different choices of $\omega$, are independent, so that these quantities would correspond to an infinite dimensional set of possibilities.

\end{remark}

\begin{remark}\label{r5}

If Schr\"odinger had realized that the pre-existing values that he had established assuming locality were impossible, as we indicated earlier in this paper, he would have proven that quantum mechanics is indeed nonlocal, using whatever observables he wished to use for the contradiction.
We note, however, that one can in fact also derive a contradiction from the pre-existing values of the original EPR variables using a Theorem of Robert Clifton \cite{RC} instead of Theorem \ref{2}, see \cite{2}. 

\end{remark}

\section{Nonlocality and Bohmian mechanics}\label{sec7}
 
In Bohmian mechanics, or pilot-wave theory, the complete state of a closed physical system composed of $N$ particles is a pair 
$(|\Psi \rangle, \bf X)$, where $|\Psi \rangle$ is the usual quantum state, and 
${\bf X}= (X_1,\ldots, X_N)$ is the configuration representing the  positions of the particles, that exist, independently of whether one measures them or not (each $X_i\in \R^3$). 

These positions are the  ``hidden variables" of the theory, in the sense that they are not included in the purely quantum description $ |\Psi \rangle$. However, they are not at all hidden: it is only the particles' positions that one detects directly, in any experiment (think, for example, of the impacts on the screen in the two-slit experiment).
Both objects, the quantum state and the particles' positions, evolve according to deterministic laws, the quantum state guiding the motion of the particles (for simplicity, we consider spinless particles):

\par\noindent
1.  The wave function evolves according to the usual Schr\"odinger's equation:
 \be
i \frac{\partial \Psi (x_1, \dots, x_N, t)}{\partial t}= (-\sum_{i=1}^N  \frac{1}{2m_i}
 \nabla_i^2+ V (x_1, \dots, x_N))\Psi (x_1, \dots, x_N, t)
 \label{Sch}
 \ee
 (with $\hbar=1$ and $m_i$ the particle's masses.)
\bigskip 

\par\noindent
2.
The particle positions ${\bf X}={\bf X}
(t)$  evolve in time, 
with their velocity being a function of the wave function. If one writes\footnote{We use lower case letters for the generic arguments of the wave function and upper case ones for the actual positions of the particles.}:
$$\Psi (x_1, \dots, x_N, t)=R (x_1, \dots, x_N, t)e^{iS (x_1, \dots, x_N, t)} 
$$ (with $R\geq 0$), then:

 \be
 \frac{ d X_k (t)}{dt}=   \displaystyle \frac{ \nabla_k S (X_1(t),\ldots,X_N(t))}{m_k},
 \label{P0}
 \ee
where $ \nabla_k$ is the gradient with respect the coordinates of the  $k$-th particle.

We will not discuss how this theory works and reproduces the quantum predictions, which in fact it does, but we will sketch the answer to the following questions (for elementary introductions to this theory, see \cite{Bri2, Tu} and for more advanced ones, see \cite{B, Bo1, BH, Bri1, DGZ, DT,  DGZ1, Go, No}):

\begin{itemize}
\item[1] 
Since Bohmian mechanics  is deterministic, the result of any experiment or measurement must be predetermined by the initial conditions of the system (possibly including those of the apparatus). But shouldn't that allow us to construct a non-contextual value map, whose mere existence is ruled out by Theorem \ref{2}?
\item[2] 
How does nonlocality manifests itself in Bohmian mechanics?
\end{itemize}

To answer the first question, one must analyze what ``measurements" really are in Bohmian mechanics. The latter introduces, besides the wave function,  only particles and their trajectories. In particular, it does not introduce pre-existing values of the spin, for example, or of any other quantum ``observable."

What are called ``measurements" in ordinary quantum mechanics are, in Bohmian mechanics, certain interactions between a particle and a measuring device. These interactions will affect the trajectories of the particles.   But, for whatever ``observable" we claim to ``measure," at the end of the experiment one detects particles' positions: a particle goes  ``up" or ``down" in a Stern-Gerlach experiment for example, or is detected at some distance from its starting point after a given time in a momentum measurement. The statistics of those positions will, according to Bohmian mechanics,  agree with the quantum predictions.

Moreover, one can show that all these measurements are actually contextual, in the sense that the result will not depend only on the initial configuration and quantum state of the measured system, but also on the way the measuring device is setup. 
For details of how this contextuality manifests itself in the measurements of spin, see \cite{1} and, for the measurement of momentum, see \cite{2}.

That is why we put here ``measurements" of quantum observables in quotation marks: they do not reveal any pre-existing value of the observables that does not depend upon how they are ``measured."
% and, indeed, the use of that word has been one of the main obstacle in understanding what goes on in quantum mechanics.
That is also why Bohmian mechanics {\it does not} provide a non-contextual value map, and is therefore not contradicted by Theorem \ref{2}.

But then, of course, Bohmian mechanics must be nonlocal. Its nonlocality follows from the fact that, in the guiding equation (\ref{P0}), the right hand side is a function of the instantaneous positions of all the particles of the system. Thus, if one affects the wave function by acting on it (by changing the potential in (\ref{Sch})) far away from where a given particle is, that change will affect instantaneously the motion of that given particle through (\ref{P0}).

To be more concrete, consider the entangled state (\ref{1}) and assume that the wave function of particle $1$ is localized in a region $A$, while that of particle $2$ is localized in region $B$, both regions being situated far apart from each other. Then introducing a potential, for example a magnetic field in a Stern-Gerlach apparatus, in region  $A$, will affect the state (\ref{1}), but, because the latter is entangled, it will also affect instantaneously  the motion of particle $2$
through equation (\ref{P0}) applied to $X_2(t)$.

\section{Conclusions}\label{sec8}
 
The fact that Bohmian mechanics is nonlocal is a virtue rather than a defect of that theory, since any physical theory reproducing some elementary quantum predictions (the perfect correlations of Section  \ref{sec2}) must be nonlocal.

Returning to what had puzzled EPR and Schr\"odinger, they thought that quantum mechanics was incomplete and that, in order to obtain a complete physical description, the description through the quantum state had to be supplemented by what we call a non-contextual value map. 
But that is because they believed that nonlocality was unthinkable. 

Later investigations by Bell \cite{Be1} and by Kochen-Specker \cite{KS} showed that such  a non-contextual value map could not exist. Meanwhile,  Bell had also shown that nonlocality was unavoidable in any physical theory reproducing some quantum predictions concerning correlations between measurements made on distant systems  \cite{Be2}. 

But, maybe more importantly, Bohm \cite{Bo1} had shown how to supplement the ordinary quantum description by introducing particle trajectories  in a manner that happens to yield a nonlocal theory that (thus) does not imply the existence of a non-contextual value map. The upshot is that quantum mechanics can indeed be completed, as EPR and Sch\"odinger thought, but not in the way they expected: not by introducing a non-contextual value map, but by giving up their assumption of locality.

 \end{document}